\documentclass[
    ,draft            % use draft while you are working on the paper
    ,numberedheadings % uncomment this option for numbered sections
  ]{aipproc}
\usepackage{amssymb}
\usepackage{enumerate}
\usepackage{graphicx}

\layoutstyle{8x11single}

\def\be{\begin{equation}}
\def\ee{\end{equation}}
\def\bea{\begin{eqnarray}}
\def\eea{\end{eqnarray}}
\def\d{\mathrm{d}}

\def\ra{\rightarrow}
\def\bh{black-hole }
\def\ns{naked-singularity }
\def\cc{cosmological constant}
\def\Kds{Kerr--de~Sitter }
\def\Sds{Schwarzschild--de~Sitter }
\let\rho\varrho
\begin{document}

\title{Basic properties of toroidal structures in Kerr--de Sitter backgrounds} 

\classification{04.70.Bw, 04.20.Dw, 95.30.Sf, 98.62.Mw, 98.80.Es}
\keywords      {cosmological constant, black holes, naked singularities,
                toroidal structures}

\author{Zden\v{e}k Stuchl\'{\i}k}{
  address={Institute of Physics, Faculty of Philosophy and Science, Silesian
  University in Opava, \\
  Bezru\v{c}ovo n\'{a}m. 13, CZ-74601 Opava, Czech Republic}
}

\author{Petr Slan\'{y}}{
%  address={}
}

\begin{abstract}
Perfect fluid tori with uniform distribution of the specific angular momentum
orbiting the Kerr--de~Sitter black holes or naked singularities are
studied. Closed equipotential surfaces corresponding to stationary toroidal
discs are allowed only in the spacetimes admitting stable circular geodesics. 
The last closed surface crosses itself in the cusp(s) enabling outflow(s) of
matter from the torus due to the violation of hydrostatic equilibrium. The
repulsive cosmological constant, $\Lambda >0$, implies the existence of the
outer cusp (with a stabilizing effect on the tori because of {\em excretion},
i.e., outflow of matter from the torus into the outer space) and the strong
collimation of open equipotential surfaces along the rotational axis. Both the
effects take place nearby so-called static radius where the gravitational
attraction is just balanced by the cosmic repulsion. The plus-family discs
(which are always corotating in the \bh backgrounds but can be
counterrotating, even with negative energy of the fluid elements, in some \ns
backgrounds) are thicker and more extended than the minus-family ones (which
are always counterrotating in all backgrounds). If parameters of the
naked-singularity spacetimes are very close to the parameters of extreme
black-hole spacetimes, the family of possible disc-like configurations
includes members with two isolated discs where the inner one is always a
counterrotating accretion disc. Mass estimates for tori with nonrelativistic
adiabatic equation of state give limits on their central mass-density, for
which the approximation of test fluid is adequate.
\end{abstract}

\maketitle

%%%%%%%%%%%%%%%%%%%%%%%%%%%%%%%%%%%%%%%%%%%%
%% MAINMATTER
%%%%%%%%%%%%%%%%%%%%%%%%%%%%%%%%%%%%%%%%%%%%

\section{Introduction}

Disc-like structures orbiting black holes seem to play an
important role in a wide range of astrophysical phenomena including the most
energetic processes in the Universe connected with quasars, or the formation
of probably the largest disc structures at all--galactic discs.
On the other hand, various cosmological observations indicate convincingly
that in the framework of inflationary cosmology a non-zero, 
although very small, vacuum energy density, i.e., a~relic repulsive
cosmological constant, $\Lambda > 0$, or some similarly acting new kind of
fields called {\em quintessence}, has to be invoked in order to explain the
dynamics of the recent Universe \cite{Bah-etal:1999:SCIEN:,
  Kol-Tur:1990:EarUni:}. Both possibilities are often referred as a {\em dark
  energy} in the Universe. It is well known that the repulsive cosmological
constant strongly influences expansion of the Universe, leading finally to an
exponentially accelerated stage. The paper shows, surprisingly enough, that the
repulsive cosmological constant could be relevant also in astrophysical
processes like the disc accretion onto a supermassive black hole and the
formation of the largest disc structures in the Universe.

Recent data from a wide variety of independent cosmological tests give the
current value of the vacuum energy density to be
\cite{Spe-etal:2003:ASTJS:} 
\be                                                          \label{e1}
      \varrho_{\rm vac(0)}\approx 0.73 \varrho_{\rm crit(0)},
\ee
where the present value of the critical energy density
$\varrho_{\rm crit(0)}$ is related with the Hubble parameter $H_{0}$
by\footnote{Geometric units $c=G=1$ are used hereafter.}
\be                                                          \label{e2}
      \varrho_{\rm crit(0)}=\frac{3H_{0}^2}{8\pi},\quad H_{0}=100h\
      \rm{km}\ \rm{s}^{-1}\ \rm{Mpc}^{-1}.
\ee
Taking value of the dimensionless parameter $h\approx 0.7$, we obtain for the
relic repulsive cosmological constant its current value
\be                                                          \label{e3}
      \Lambda_{0}=8\pi\varrho_{\mathrm{vac(0)}}\approx 1.3 \times 10^{-56}\
      \rm{cm}^{-2}.
\ee

Basic properties of geometrically thin (accretion) discs with low accretion
rates and negligible pressure are given by the circular geodesic motion in the
black-hole backgrounds \cite{Nov-Tho:1973:BlaHol:}, while for geometrically
thick discs with high accretion rates and pressure being relevant
they are determined by equipotential surfaces of test perfect fluid rotating
in the background (see, e.g., \cite{Abr:1998:TheoryBlackHoleAccretionDisks:}).
The equipotential surfaces can be closed or open. Moreover, there is a special
class of critical surfaces self-crossing in a cusp(s), which can be either
marginally closed or open. The closed---toroidal---equipotential surfaces
determine stationary configurations (tori). The fluid can fill any toroidal
equipotential surface; at the surface of the torus pressure vanishes, 
but its gradient is non-zero \cite{Koz-Jar-Abr:1978:ASTRA:}. On the other
hand, the open equipotential surfaces are important in dynamical situations,
e.g., in modeling of jets \cite{LyB:1969:NATURE:, Bla:1987:300YoG:}, as the
matter can flow along the open surfaces. The
critical---marginally closed---equipotential surfaces $W_{\rm crit}$ are
important in the theory of thick accretion discs, because accretion onto the
black hole through a cusp of the equipotential surface, located in the
equatorial plane, is possible due to a small overcoming of the critical 
surface by the surface of the disc (Paczy\'nski mechanism). Accretion is thus
driven by a violation of the hydrostatic equilibrium, rather than by viscosity
of the accreting matter \cite{Koz-Jar-Abr:1978:ASTRA:}.

The equatorial circular motion of test particles (Keplerian motion) and its
relevance for thin accretion discs were studied thoroughly in the case of \Sds
(SdS) geometry in \cite{Stu-Hle:1999:PHYSR4:}, and in the case of \Kds (KdS)
geometry in \cite{Stu-Sla:2004:PHYSR4:}, even if some
important remarks can also be found in the work of Carter or Demianski
\cite{Car:1973:BlaHol:, Dem:1973:ACTAAS:}. It was shown that thin discs have,
besides the standard inner edge at the inner marginally stable circular orbit,
an outer edge at the outer marginally stable circular orbit located slightly
under the static radius of the given spacetime. The static radius is an
unstable 
circular orbit where the gravitational attraction of the black hole is just
compensated by the cosmic repulsion and where the static geodesic observer,
i.e., the geodesic observer with only the time-component of its 4-velocity
being non-zero, resides. Moreover the accretion efficiency, $\eta=E_{\rm
  ms(o)}-E_{\rm ms(i)}$, given by the difference of the specific energies at
the outer and the inner marginally stable orbits, is smaller when compared
with the asymptotically flat geometries, especially due to the fact that
$E_{\rm ms(o)}<1$. 

Equilibrium toroidal configurations of barotropic perfect fluid orbiting in
the \Sds background were studied in \cite{Stu-Sla-Hle:2000:ASTRA:}. An
analysis of the structure of equipotential surfaces revealed the existence of
another (outer) critical point (the so-called cusp) in which one of the
equipotential surfaces is self-crossing. As the discussed equilibrium of fluid
tori is the hydrostatic equilibrium, the outer cusp, when it is located on the
marginally 
closed equipotential surface, determines the outer edge of the disc, exactly in
the same way as the inner cusp determines the inner edge
\cite{Koz-Jar-Abr:1978:ASTRA:}. 
Moreover, the outflows through the outer cusp can
stabilize the accretion discs against so-called \textit{runaway instability}
\cite{Abr-Cal-Nob:1983:NATURE:},
as shown by Rezzolla et al. \cite{Rez-Zan-Fon:2003:ASTRA:}. When
the critical surface with the inner cusp is open or does not exist, while the
critical surface with the outer cusp is marginally closed, such a
configuration corresponds to a completely new type of discs called
\emph{excretion disc}. More detailed description of possible
equilibrium configurations of barotropic fluid in the backgrounds with
$\Lambda>0$ is given in the Sec. \ref{s3} below where the situation is
discussed in the case of KdS geometry. Strong collimation of open
equipotential surfaces near the axis of rotation suggests a tendency of
backgrounds with $\Lambda>0$ to collimate streams of particles (jets) moving
along the rotational axis.

Equilibrium toroidal configurations of barotropic perfect fluid orbiting in
the \Kds background were analyzed thoroughly in \cite{Sla-Stu:2005:CLAQG:}
where also the case of naked singularities was discussed. Here we present an
overview of the basic results obtained in the approximation of test discs, and
add an estimation for the central mass-energy density of an adiabatic fluid
for which the obtained test-fluid results are valid.

\section{\Kds geometry} \label{s2}

The geometry of \Kds spacetimes is given by the line element
\be
 \d s^2 = -\frac{\Delta_r}{I^2 \rho^2}(\d t-a\sin^2 \theta \d\phi)^2
          +\frac{\Delta_{\theta}\sin^2 \theta}{I^2 \rho^2}
               \left[a\d t- \left(r^2+a^2 \right) \d\phi \right]^2 + 
               \frac{\rho^2}{\Delta_r} \d r^2 +
               \frac{\rho^2}{\Delta_{\theta}} \d\theta^2     
\label{e10}
\ee
where
\be
    \Delta_r = r^2 -2Mr + a^2 -\frac{1}{3}\Lambda r^2 
                   \left(r^2+a^2 \right), \quad
    \Delta_{\theta} = 1+ \frac{1}{3} \Lambda a^2 \cos^2 \theta, \quad
    I = 1+ \frac{1}{3} \Lambda a^2, \quad \rho^2 = r^2 +a^2 \cos^2 \theta.
                                                                \label{e14}
\ee
The spacetime is specified by three parameters: central mass $(M)$, rotational
parameter $(a)$ corresponding to the specific angular momentum of the central
object, and positive cosmological constant $(\Lambda)$. It is convenient to
introduce the dimensionless ``cosmological parameter''
\be
     y = \frac{1}{3} \Lambda M^2
\label{e15}
\ee
and reformulate relations (\ref{e10})--(\ref{e14}) into the completely
dimensionless form by putting $M=1$ hereafter. The spacetime is stationary,
axially symmetric and asymptotically de Sitter. 

The spacetime horizons are determined by the condition $\Delta_r=0$. In
general, three horizons (the inner and the outer of the black hole, $r_{\rm
  h-}$ and $r_{\rm h+}$, and the cosmological one, $r_{\rm c}$) exist. A more
detailed discussion on the existence of event horizons of KdS geometry
including the extreme cases in which two or even three horizons coincide can
be found in \cite{Stu-Sla:2004:PHYSR4:}. Note that cosmological horizon, behind
which the geometry is dynamic, exists in all KdS spacetimes, and that there is
a maximal value of the cosmological parameter allowing the existence of 
black holes: $y_{\rm crit}\doteq 0.059$.

\begin{figure}
\centering
\includegraphics[width=.5 \hsize]{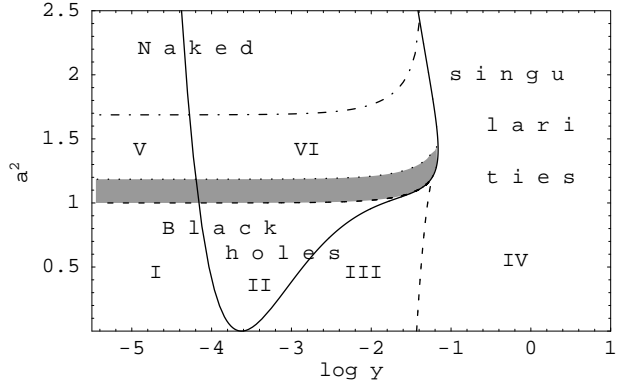}
\caption{Division of KdS spacetimes according to the existence of stable
  equatorial circular orbits (SECOs). (\textbf{I}) Black-hole (BH) spacetimes
  with both direct and retrograde SECOs. (\textbf{II}) BH spacetimes with no
  retrograde SECOs. (\textbf{III}) BH spacetimes with no SECOs. (\textbf{IV})
  Naked-singularity (NS) spacetimes with no SECOs. (\textbf{V}) NS spacetimes
  with both direct and retrograde SECOs. (\textbf{VI}) NS spacetimes with no
  retrograde SECOs. The dashed-dotted curve forms the boundary of NS region,
  where the retrograde plus-family SECOs exist; shaded is its subregion where
  such SECOs possess negative energy.}
\label{f1}
\end{figure}

An analysis of the equatorial circular orbits of test particles in terms of the
stability against radial perturbations enables to divide the parametric space
$(y,\, a^2)$ into six regions corresponding to the black-hole/\ns spacetimes
containing stable equatorial circular geodesics of a given family
(Fig. \ref{f1}). In the \bh backgrounds, the plus-family corresponds to the
stable corotating orbits while the minus-family to the stable counterrotating
ones, where the direction of the orbits is related to the locally non-rotating
frames (LNRFs) introduced by Bardeen \textit{et al.}
\cite{Bar-Pre-Teu:1972:ASTRJ2:}; 
the orbits with locally measured azimuthal component of its 
4-momentum $P^{(\phi)}>0$ are called direct or corotating, while the ones with
$P^{(\phi)}<0$ are called retrograde or counterrotating. In the \ns
backgrounds, there is a sort of stable plus-family orbits which are
counterrotating from the point of view of LNRFs, even with negative specific
energy (the shaded region in Fig. \ref{f1}). 
Note that also in \bh backgrounds, contrary to the Kerr case ($y=0$),
the plus-family orbits can be both direct or retrograde, however, the
retrograde ones, located near the upper limiting radius for the existence of
the plus-family orbits (static radius or the retrograde photon circular orbit),
are unstable.
Moreover, for each family of orbits there is a maximal value of
the cosmological parameter $y$ allowing the stable circular orbits of a given
family: $y_{\rm c(ms+)}\doteq 0.069$ and $y_{\rm c(ms-)}=12/15^4
\doteq 0.00024$; the latter coincides with the maximal value of $y$ allowing
the stable circular orbits in SdS spacetimes \cite{Stu-Hle:1999:PHYSR4:}.
A detailed discussion of the direction and stability of equatorial circular
orbits can be found in \cite{Stu-Sla:2004:PHYSR4:}.

%%%%%%%%%%%%%%%%%%%%%%%%%%%%%%%% Fig 1 %%%%%%%%%%%%%%%%%%%%%%%%%%%%%%%%%%%%%%%
%%%%%%%%%%%%%%%%%%%%%%%%%%%%%%%%%%%%%%%%%%%%%%%%%%%%%%%%%%%%%%%%%%%%%%%%%%%%%%

\section{Equipotential surfaces} \label{s3}

Analytic theory of
equilibrium configurations of rotating perfect fluid bodies was developed by
Boyer \cite{Boy:1965:PCPS:} and than studied by many authors. The main result
of the theory, known as ``Boyer's condition'', states that the boundary of any
stationary, barotropic, perfect fluid body has to be an equipotential
surface. Here we shall summarize its application to the relativistic test
perfect fluid orbiting in a stationary and axisymmetric way in a stationary,
axisymmetric background, as was introduced by Abramowicz and co-workers in the
case of Schwarzschild and Kerr black holes
\cite{Koz-Jar-Abr:1978:ASTRA:, Abr-Jar-Sik:1978:ASTRA:}, and use it for an
analysis in the KdS spacetimes \cite{Sla-Stu:2005:CLAQG:}. 

In the standard Boyer--Lindquist coordinates the spacetime is described by the
line element 
\be                                                          \label{e34}
     \d s^2=g_{tt}\d t^2 + 2g_{t\phi}\d t \d\phi +
     g_{\phi\phi}\d\phi^2 + g_{rr}\d r^2 + g_{\theta\theta}\d\theta^2
\ee
satisfying the properties of stationarity and axial symmetry, i.e.,
$\partial_t g_{\mu\nu}=\partial_{\phi}g_{\mu\nu}=0$. The stress-energy tensor
of~perfect fluid is given by 
\be                                                          \label{e35}
     T^\mu_{\hphantom{\mu}\nu} = (\epsilon +p) U^\mu U_\nu + p\,\delta^\mu_\nu,
\ee
where $\epsilon$  and~$p$ are total energy density and pressure of~the
fluid, respectively. Further, we consider test perfect fluid with the
4-velocity $U^{\mu}=(U^{t},\, U^{\phi},\, 0,\, 0)$; orbital motion of the
fluid is
characterized by vector fields of the angular velocity $\Omega \left(
  r,\,\theta \right)=U^{\phi}/U^{t}$ and~the specific angular momentum 
$\ell \left( r,\, \theta \right)= -U_{\phi}/U_{t}$,
related by the metric coefficients of the background
\be                                                          \label{e38}
     \Omega = - \frac{g_{t\phi} + \ell g_{tt}}{g_{\phi\phi} + \ell g_{t\phi}}.
\ee
The relativistic Euler equation in the axially symmetric form reads
\be                                                          \label{e39}
     \frac{\partial_{i} p}{\epsilon +p} = -\partial_{i} (\ln U_t) +
     \frac{\Omega\,\partial_{i} \ell}{1- \Omega \ell},
\ee
where $i=r,\ \theta$ and
\be                                                          \label{e40}
     (U_t)^2 = \frac{g^2_{t\phi} - g_{tt}\,g_{\phi\phi}}
               {g_{\phi\phi} + 2 \ell g_{t\phi} + \ell^2 g_{tt}}.
\ee

For a barotropic fluid, i.e., for a body with an equation of state
$p=p(\epsilon)$, the surfaces of constant pressure are given
by the equipotential surfaces of the potential
$W(r,\,\theta)$ defined by the relations 
\cite{Abr-Jar-Sik:1978:ASTRA:}
\be                                                          \label{e41}
     \int_{0}^{p}\frac{\d p}{\epsilon + p} = 
     \ln(U_t)_{\rm in}-\ln(U_t)+\int_{\ell_{\rm in}}^{\ell}\frac{\Omega\d
     \ell}{1-\Omega \ell} \equiv W_{\rm in} - W, 
\ee
where the subscript ``in'' refers to the inner edge of the disc. The explicit
form of the potential, $W=W(r,\,\theta)$, is given by the relations
(\ref{e40}) and (\ref{e41}), if one
specifies the metric tensor of the background and the ``rotational law'',
i.e., the function $\Omega=\Omega(l)$. The simplest but astrophysically
very important is the case of uniform distribution of the specific angular
momentum,
\be                                                          \label{e42}
     \ell(r,\,\theta)=\mbox{const},
\ee
through the disc.
It is well known that the tori with $\ell(r,\,\theta)=\mbox{const}$
are marginally stable \cite{Seg:1975:ASTRJ2:} and capable to produce maximal
luminosity at all \cite{Abr-Cal-Nob:1980:ASTRJ2:}. Note that topological
properties of the equipotential surfaces seem to be rather independent of the
distribution of the specific angular momentum $\ell(r,\,\theta)$, see, e.g.,
\cite{Abr-Cal-Nob:1980:ASTRJ2:, Jar-Abr-Pac:1980:ACTAS:,
  Abr:1998:TheoryBlackHoleAccretionDisks:}. In this special case the potential
is given by very simple formula 
\be                                                          \label{e43}
     W(r,\, \theta)=\ln U_{t}(r,\, \theta).
\ee  
The
points where $\partial_i W = 0$ correspond to free-particle (geodesic) motion
due to the vanishing of the pressure-gradient forces there. Moreover, at the
center of any perfect fluid torus the pressure attains the extreme value
(maximum) and matter must follow a stable geodesic there. 
Thus, thick discs can exist only in the backgrounds allowing the motion along
stable circular geodetical orbits.

In the KdS backgrounds, the potential (\ref{e43}) takes the form
\be                                                          \label{e46}
     W(r,\,\theta) = \ln\left\{\frac{\rho^2}{I^2}\frac{\Delta_r
          \Delta_{\theta} \sin^2 \theta}{\Delta_{\theta}(r^2+a^2-a\ell)^2 
          \sin^2 \theta - \Delta_r (\ell-a\sin^2 \theta)^2}\right\}^{1/2}.
\ee
Performing the limit $a\ra 0$, we get the corresponding form of the potential
(\ref{e43}) in the SdS backgrounds \cite{Stu-Sla-Hle:2000:ASTRA:}. All
relevant properties of the 
equipotential surfaces are determined by the behaviour of the potential in the
equatorial plane ($\theta=\pi/2$). Inspecting the reality conditions of
$W(r,\,\theta=\pi/2)$ in the stationary parts of the background ($\Delta_r
\geq 0$; in the \bh backgrounds, we restrict our attention to the
stationary region between the (outer) \bh and cosmological horizon only), we
arrive at the condition for the occurrence of matter with a given distribution
of $\ell(r,\,\theta)$,
\be                                                          \label{e48}
     \ell_{\rm ph-}<\ell<\ell_{\rm ph+},
\ee 
where
\be                                                          \label{e49}
     \ell_{\rm ph\pm}(r;\,a,\,y) = a + \frac{r^2}{a\pm\sqrt{\Delta_r}} 
\ee
corresponds to the effective potential of the photon geodesic motion;
see \cite{Stu-Hle:2000:CLAQG:} for an alternative definition. Local extrema of
the function $W(r,\,\theta=\pi/2)$ lie at those radii where the specific
angular momentum coincides with the specific angular momentum of test
particles moving on the geodetical (Keplerian) circular orbits, i.e., where
\be                                                          \label{e50}
     \ell=\ell_{\rm K\pm}(r;\,a,\,y)\equiv \pm\frac{(r^2+a^2)(1-yr^3)^{1/2} \mp
     ar^{1/2}[2+r(r^2+a^2)y]}{r^{3/2}[1-(r^2+a^2)y]-2r^{1/2} \pm 
     a(1-yr^3)^{1/2}}. 
\ee
Those extrema are the only local extrema of the function $W(r,\,\theta)$. 

Toroidal configurations arise for such a distribution of $\ell(r,\,\theta)$ in
the disc which intersects the Keplerian distribution $\ell_{\rm K\pm}(r)$ in
the part(s) corresponding to stable circular orbits. In \bh backgrounds,
stationary toroidal configurations exist for $\ell\in (\ell_{\rm ms(i)},
\,\ell_{\rm ms(o)})$, where $\ell_{\rm ms(i)}\ (\ell_{\rm ms(o)})$ corresponds
to the Keplerian specific angular momentum on the inner (outer) marginally
stable orbit. The same is true also in most of the \ns 
backgrounds, however, exceptions exist concerning the plus-family discs in \ns
backgrounds with the rotational parameter low enough to admit counterrotating
stable plus-family circular geodesics, see Fig. \ref{f1}. In fact, there are
\ns backgrounds, in which the stationary tori exist for any uniform
distribution of $\ell(r,\,\theta)$ in the disc \cite{Sla-Stu:2005:CLAQG:}. 

%%%%%%%%%%%%%%%%%%%%%%%%%%%%%%%% Fig 2 %%%%%%%%%%%%%%%%%%%%%%%%%%%%%%%%%%%%%%%
\begin{figure}
\begin{minipage}{.32\linewidth}
\centering
\begin{picture}(47,48)
\put(0,0){\includegraphics[width=1 \hsize]{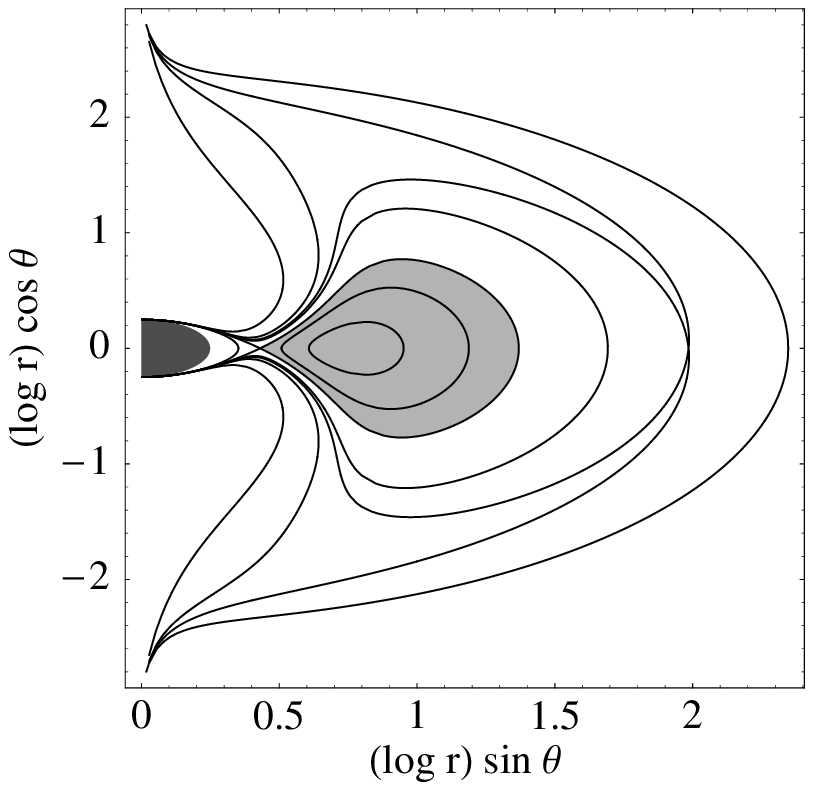}}
\put(27,48){\small{$W_{\rm crit(i)}<W_{\rm crit(o)}$}}
\put(27,10){\small{$\ell_{\rm ms(i)}<\ell<\ell_{\rm mb}$}}
\end{picture}
\par\small (a)\enspace
\end{minipage} \hfill %
\begin{minipage}{.32\linewidth}
\centering
\begin{picture}(47,48)
\put(0,0){\includegraphics[width=1 \hsize]{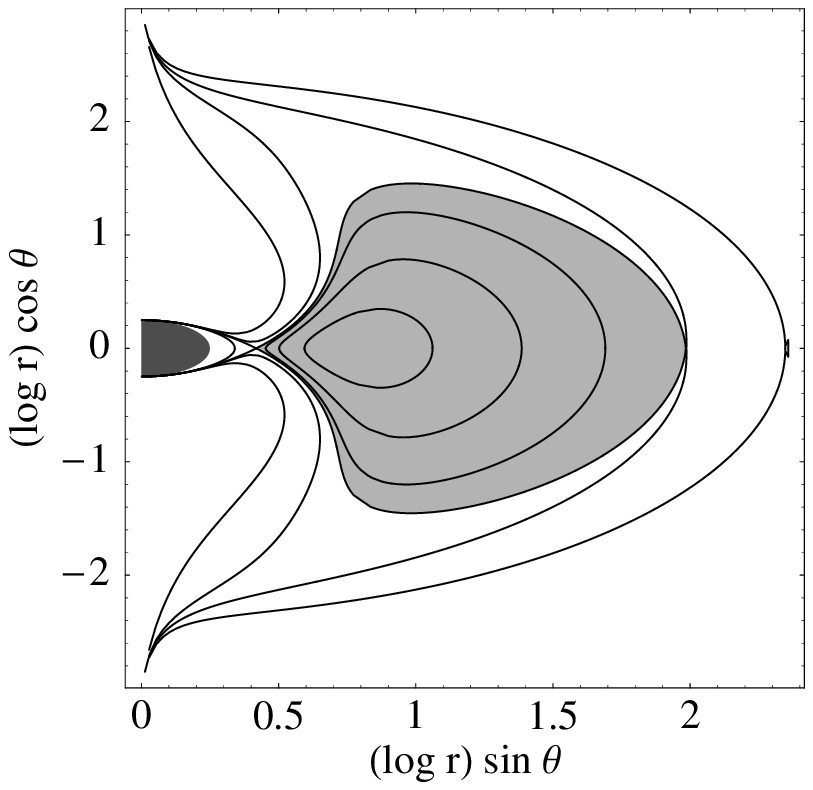}}
\put(27,48){\small{$W_{\rm crit(i)}=W_{\rm crit(o)}$}}
\put(38,10){\small{$\ell=\ell_{\rm mb}$}}
\end{picture}
\par\small (b)\enspace
\end{minipage} \hfill %
\begin{minipage}{.32\linewidth}
\centering
\begin{picture}(47,47)
\put(0,0){\includegraphics[width=1 \hsize]{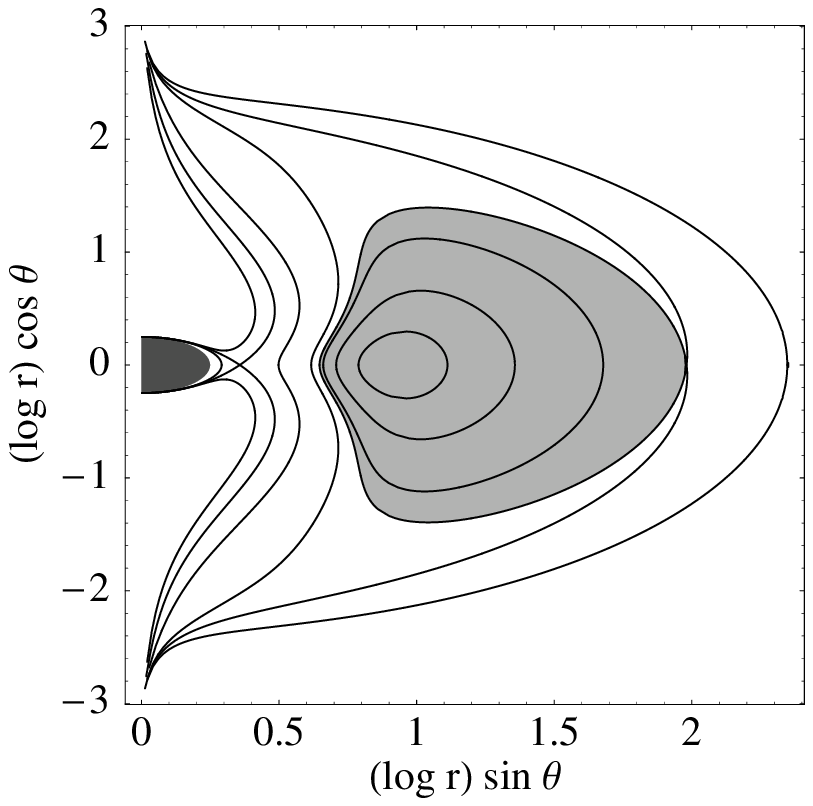}}
\put(27,47.5){\small{$W_{\rm crit(i)}>W_{\rm crit(o)}$}}
\put(27,9.5){\small{$\ell_{\rm mb}<\ell<\ell_{\rm ms(o)}$}}
\end{picture}
\par\small (c)\enspace
\end{minipage}
\caption{Typical behaviour of equipotential surfaces (meridional sections) in
  SdS and KdS \bh spacetimes. Light gray region contains closed
  equipotential surfaces. The last closed surface is self-crossing in the
  cusp(s). Possible toroidal configurations correspond to: (a) accretion
  discs, (b) marginally bound accretion discs and (c) excretion discs. (The
  figures are drawn for the KdS spacetime with $y=10^{-6}$, $a^2=0.6$ using
  the standard Boyer-Linquist coordinates.)}
\label{f2}
\end{figure}
%%%%%%%%%%%%%%%%%%%%%%%%%%%%%%%%%%%%%%%%%%%%%%%%%%%%%%%%%%%%%%%%%%%%%%%%%%%%%%

In both \bh and \ns backgrounds we can distinguish three kinds of discs
(Figure~\ref{f2}):
\begin{description}
\item [accretion discs:] 
  Toroidal equipotential surfaces are bounded by the marginally closed
  critical equipotential surface self-crossing in the inner cusp and enabling
  outflow of matter from the disc into the BH/NS. Another critical surface
  self-crossing in the outer cusp is open. Matter filling the region between
  the critical surfaces cannot remain in hydrostatic equilibrium and
  contributes to the accretion flow along the inner cusp and a throat formed
  by open surfaces. Moreover, if the potential levels corresponding to the
  critical surfaces are comparable, i.e., $W_{\rm crit(i)} \lesssim W_{\rm
  crit(o)}$, huge overfilling of the critical surface with the inner cusp
  causing the accretion could be combined with the so-called {\em excretion},
  i.e., outflow through the outer cusp (after overfilling of the critical
  surface with the outer cusp), having a capability to regulate the accretion. 
\item [marginally bound accretion discs:] 
  Such configurations exist only for the uniform distribution of the specific
  angular momentum in the disc $\ell(r,\,\theta)=\ell_{\rm mb}$, where
  $\ell_{\rm mb}$ corresponds to the Keplerian specific angular momentum on
  the marginally bound circular orbit.
  Toroidal equipotential surfaces are bounded by the marginally closed
  critical equipotential surface self-crossing in both the cusps. Any
  overfilling of the critical surface causes the accretion inflow through the
  inner cusp as well as the excretion outflow through the outer cusp.
\item [excretion discs:]
  Toroidal equipotential surfaces are bounded by the marginally closed
  critical equipotential surface self-crossing in the outer cusp and enabling
  outflow of matter from the disc into the outer space by a violation of
  hydrostatic equilibrium. The equipotential surface with the inner cusp, if
  such a surface exists, is open (cylindrical) and separated from the critical
  surface with the outer cusp by other cylindrical surfaces which, in fact,
  disable accretion into the black hole.
\end{description}

\begin{figure}
\begin{minipage}{.32\linewidth}
\centering
\includegraphics[width=1 \hsize]{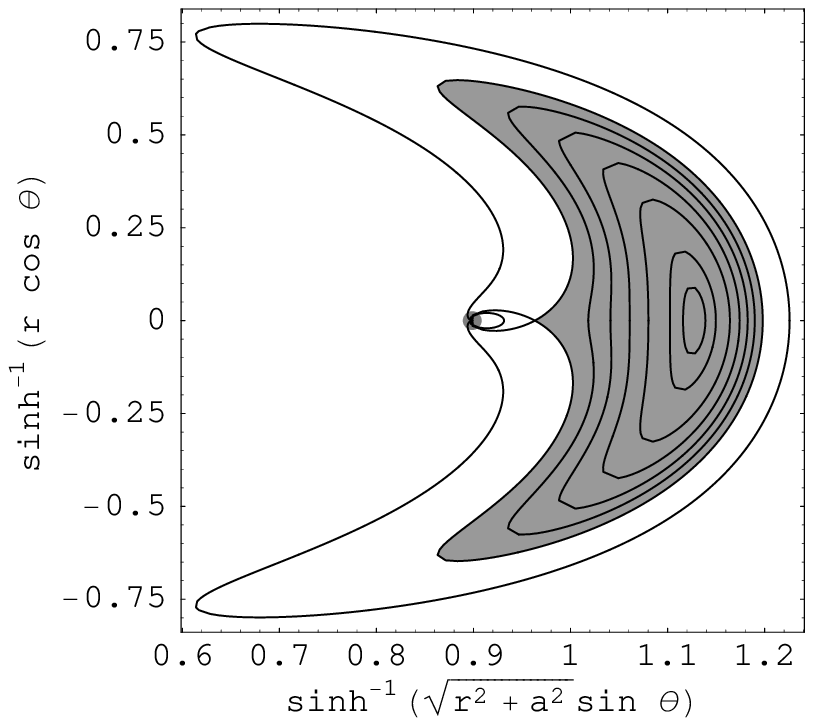}
\par\small (a)\enspace
\end{minipage} \hfill %
\begin{minipage}{.32\linewidth}
\centering
\includegraphics[width=.92 \hsize]{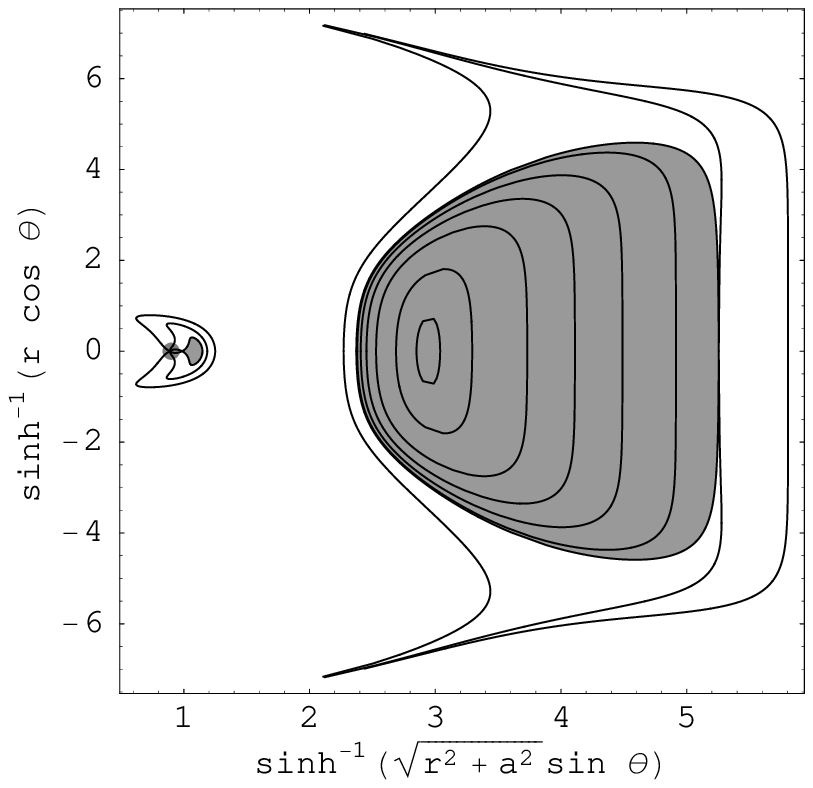}
\par\small (b)\enspace
\end{minipage} \hfill %
\begin{minipage}{.32\linewidth}
\centering
\includegraphics[width=.92 \hsize]{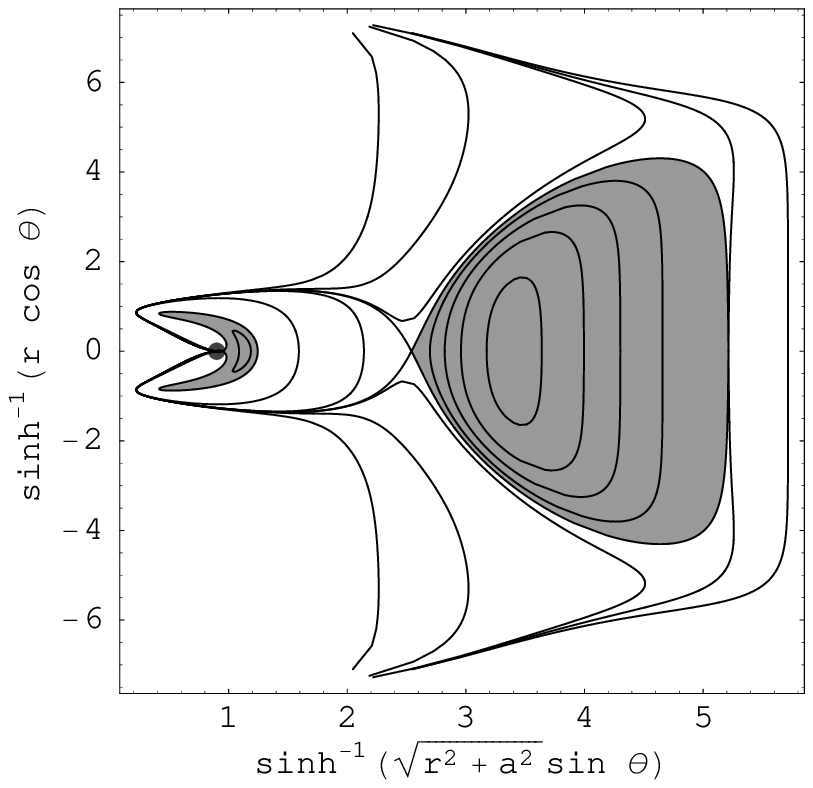}
\par\small (c)\enspace
\end{minipage}
\caption{Examples of ``exotic'' toroidal configurations (meridional sections
  through the equipotential surfaces) in KdS \ns spacetimes. Shaded regions
  contain closed equipotential surfaces. The last closed surface is
  self-crossing in the cusp(s). (a) Counterrotating negative-energy accretion
  disc. (b) Two isolated discs -- the inner one is a counterrotating accretion 
  disc, the outer one is a corotating excretion disc. (c) Two isolated discs
  -- the inner one is a counterrotating accretion disc, the outer one is a
  marginally bound counterrotating accretion disc. (The figures are drawn for
  NS spacetimes with $y=10^{-6},\ a^2=1.05$ and the values of $\ell$ being
  equal to $8$, $3.5$ and $-4.72$, respectively. The Kerr--Schild coordinates,
  covering the whole range of the disc including the ring singularity, and
  their appropriate scaling are used.) }
\label{f3}
\end{figure}

In KdS \ns backgrounds allowing counterrotating stable plus-family
circular orbits (Fig. \ref{f1}), some more exotic configurations exist:
\begin{enumerate}[a.]
\item counterrotating negative-energy discs \\[.5ex]
  Specific energy of the fluid
  elements in the center and the cusps (where the fluid follows the geodesic
  motion) is negative and we can expect that every fluid element in the disc 
  has energy $E<0$. Moreover, no open equipotential surfaces going out from
  the singularity are connected with such configurations. In the case of
  accretion discs, the equipotential surface with the outer cusp need not
  exist (Fig.~\ref{f3}a). 
\item configurations with two isolated discs corresponding to the same value
  of $\ell$ \\[.5ex]
  The inner disc is always counterrotating accretion disc (with matter in
  the states with $E<0$ or $E>0$), the outer disc can be, in dependence on the
  value of $\ell$, the corotating or counterrotating excretion disc, as well as
  the counterrotating accretion disc (Figs.~\ref{f3}b, c).
\end{enumerate}

%%%%%%%%%%%%%%%%%%%%%%%%%%%%%%%% Fig 3 %%%%%%%%%%%%%%%%%%%%%%%%%%%%%%%%%%%%%%%
%%%%%%%%%%%%%%%%%%%%%%%%%%%%%%%%%%%%%%%%%%%%%%%%%%%%%%%%%%%%%%%%%%%%%%%%%%%%%%

Due to the existence of non-zero pressure-gradients in the fluid, the inner
edge of accretion
discs (corresponding to the inner cusp of equipotential surfaces) is shifted
under the inner marginally stable circular orbit up to the inner marginally
bound circular 
orbit, $r_{\rm mb(i)}<r_{\rm in}<r_{\rm ms(i)}$. Similarly, the outer edge of
excretion discs (corresponding to the outer cusp of equipotential surfaces) is
located between the outer marginally stable and outer marginally bound
circular orbit,
$r_{\rm ms(o)}<r_{\rm out}<r_{\rm mb(o)}$.
Marginally bound accretion discs have, thus, naturally determined both edges
by the location of the cusps of the only critical surface, $r_{\rm in}\approx
r_{\rm mb(i)},\ r_{\rm out}\approx r_{\rm mb(o)}$, and correspond to maximally
extended discs. Moreover, potential difference between the boundary
(determined by the marginally closed critical surface) and the center of the
torus, $\Delta W=W_{\rm crit}-W_{\rm center}$, takes the largest values for
plus-family marginally bound accretion discs. In \bh backgrounds, the
maximal value corresponds to the disc corotating the extreme Kerr black hole
($y=0$), $\Delta W\approx 0.549$ \cite{Abr-Jar-Sik:1978:ASTRA:}, and with
the cosmological parameter $y$ growing up to $y_{\rm crit}\doteq 0.059$
tends to zero. In \ns backgrounds, the potential difference grows unlimitedly,
$\Delta W \to\infty$, for the plus-family discs orbiting a naked singularity
approaching the extreme-hole state, independently of the cosmological
parameter $y<y_{\rm crit}$. 

\section{Mass estimates of adiabatic tori}\label{s4} 

The total mass of the disc, $m$, is given by the Tolman's formula
\cite{Koz-Jar-Abr:1978:ASTRA:} 
\be                                                          \label{e51}
     m=\int_{\rm disc}\left (-T^{t}_{t} + T^{r}_{r} +
       T^{\theta}_{\theta} + T^{\phi}_{\phi}\right ) \sqrt{-g}\,
     \d r\,\d\theta\,\d\phi 
\ee
where $g=\mbox{det}(g_{\mu\nu})$ and $T^{\mu}_{\nu}$ is the stress-energy
tensor of perfect fluid given by the relation (\ref{e35}). 
Assuming an adiabatic equation of state for a barotropic perfect fluid
\cite{Too:1965:ASTRJ2:} 
\be                                                          \label{e52}
     p=K\rho^{\gamma}, \quad \gamma=1+\frac{1}{n}
\ee
where $\gamma$ is an adiabatic index, 
and a non-relativistic limit with $p\ll\epsilon\approx\rho$, where $\rho$ is
the rest-mass density, 
the relation (\ref{e51}) can be written
in an approximative form \cite{Koz-Jar-Abr:1978:ASTRA:} 
\be                                                          \label{e53}
     m=2\pi\rho_{\rm c}\int_{\rm disc}\left[
       \frac{1+\ell\,\Omega(r,\,\theta)}{1-\ell\,\Omega(r,\,\theta)}\right ]
       \left[\frac{W_{\rm in}-W(r,\,\theta)}{W_{\rm in}-W_{\rm c}}\right ]^n
       \left(r^2+a^2\cos^2\theta \right )\sin\theta\,\d r\,\d\theta
\ee
in which the functions $\Omega(r,\,\theta)$ and $W(r,\,\theta)$ are given by
the relations (\ref{e38}) and (\ref{e46}). $W_{\rm c}$ and $\rho_{\rm c}$
correspond to the potential value and the rest-mass density in the centre of
the disc, respectively. 

Comparing the total mass of the disc $m$ with the mass of the black hole $M$,
i.e. $m\approx M$,
we can get the maximal value of the central mass-density for which the
approximation of test fluid is valid in the sense $m\ll M$. The first results
of numerical integration have been obtained for marginally bound accretion
discs orbiting in the SdS background and for two principal values of adiabatic
index $\gamma=5/3$ and $\gamma=7/5$. The results are presented in Table
\ref{t1}. 
 
\section{Conclusions}\label{conc}

We conclude that there are two features in the structure of equipotential
surfaces near the static radius of a given spacetime connected with a cosmic
repulsion; both have already been known from the previous analysis of
equipotential surfaces in SdS backgrounds \cite{Stu-Sla-Hle:2000:ASTRA:}. 
The first one is the outer cusp 
which enables the outflow of matter from the new type of stationary tori---the
{\em excretion discs}. 
However, the outer cusp plays an important role also for the accretion discs, 
as puts an upper limit on their extension. 
Recall that no such an outer edge is naturally defined in the
case of accretion discs orbiting a single black hole (naked singularity) in
asymptotically flat Schwarzschild or Kerr backgrounds.
The second feature connected with the cosmic repulsion consists in strong
collimation of open equipotential surfaces near the axis of rotation,
being evident nearby and behind the static radius, when compared with the Kerr
case, see \cite{Sla-Stu:2005:CLAQG:} for an illustrative figure, suggesting a
certain role of $\Lambda>0$ in the collimation of jets far away from the
maternal accretion or excretion disc. 
%%%%%%%%%%%%%%%%%%%%%%%%%%%%%%%% Fig 4 %%%%%%%%%%%%%%%%%%%%%%%%%%%%%%%%%%%%%%%
%\begin{figure}
%\begin{minipage}{.49\hsize}
%\centering
%\includegraphics[width=.45 \hsize]{jets}
%\end{minipage}\hfill%
%\begin{minipage}{.49\hsize}
%\includegraphics[width=.99 \hsize]{shape}
%\end{minipage}
%\caption{(left) Influence of a repulsive cosmological constant on the
%  structure of equipotential surfaces. Configurations corresponding to
%  accretion discs corotating the Kerr ($y=0,\ a^2=0.99$) and the KdS
%  ($y=10^{-6},\ a^2=0.99$) black hole are compared.
%%   (right) Influence of spacetime's rotation on the shape of marginally bound
%%   accretion discs ($\ell = \ell_{\rm mb}$) orbiting the KdS black hole
%%   ($y=10^{-6},\ a^2=0.99$). For comparison, thick marginally bound accretion
%%   disc orbiting the SdS black hole ($y=10^{-6},\ a=0$) is presented.
%}
%\label{f4}
%\end{figure}
%%%%%%%%%%%%%%%%%%%%%%%%%%%%%%%%%%%%%%%%%%%%%%%%%%%%%%%%%%%%%%%%%%%%%%%%%%%%%%
Rotation of the background
influences the shape of tori: the corotating discs are thicker and more
extended than the discs in corresponding SdS background, generating a narrower
funnel where the jets are most probably created
\cite{Fra-Kin-Rai:2002:AccretionPower:}. On the other hand, the same is true
for the discs in SdS backgrounds when these are compared with counterrotating
discs in corresponding KdS background.

%%%%%%%%%%%%%%%%%%%%%%%%%%%%%%%%%% table 1 %%%%%%%%%%%%%%%%%%%%%%%%%%%%%%%%%%%
\begin{table}
\centering
\caption{(Upper part) Mass parameter, the static radius and radius of the
 outer marginally stable orbit in extreme KdS black-hole
 spacetimes with the current value of the \cc, $\Lambda_0\approx 1.3
 \times 10^{-56}\ \rm{cm}^{-2}$. 
 (Lower part) Central mass-density of an adiabatic fluid for which the total
 mass of the disc is comparable with the mass of the SdS black hole for two
 values of adiabatic index $\gamma=5/3$ and $\gamma=7/5$.}
%\vskip.5ex
\begin{tabular}{llllllll}
\hline
\rule{0pt}{8pt}
 $y$ & $10^{-44}$ & $10^{-42}$ & $10^{-34}$ & $10^{-30}$ & $10^{-28}$ &
 $10^{-26}$ & $10^{-22}$ \\ 
 $M/M_{\odot}$ & 10 & 100 & $10^6$ & $10^8$ & $10^9$ & $10^{10}$ & $10^{12}$ \\
 $r_{\mathrm{s}}/\mbox{[kpc]}$ & 0.2 & 0.5 & 11 & 50 & 110 & 230 & 1100 \\
 $r_{\mathrm{ms}}/\mbox{[kpc]}$ & 0.15 & 0.3 & 6.7 & 31 & 67 & 150 & 670 \\
\hline
$\rho^{(5/3)}_{\rm c}/[{\rm kg\,m}^{-3}]$ & & & & $10^{-25}$ & $10^{-23}$ & &
\\ 
$\rho^{(7/5)}_{\rm c}/[{\rm kg\,m}^{-3}]$ & & & & $10^{-17}$ & $10^{-16}$ & &
\\ 
\hline
\end{tabular}
\label{t1}
\end{table}
%%%%%%%%%%%%%%%%%%%%%%%%%%%%%%%%%%%%%%%%%%%%%%%%%%%%%%%%%%%%%%%%%%%%%%%%%%%%%%

Finally, we shall give an idea on scales, at which the discussed effects take
place, by expressing basic characteristics of the tori in astrophysical units
for the current value of the \cc, $\Lambda=\Lambda_0$. The results are
presented in Table~\ref{t1}. As the outer edge of tori is located between the
outer marginally stable orbit and the static radius, $r_{\rm ms(o)}<r_{\rm
  out}<r_{\rm s}$, the repulsive cosmological constant puts a limit on maximal
extension of disc-like structures in a given background. Remarkably for
supermassive black holes ($10^6 M_\odot$--$10^{10}
M_\odot$), the dimensions of {\em test} tori are roughly comparable with the
dimensions of galaxies \cite{Car-Ost:1996:ModAst:}. 
The upper limits on the central mass-density of the adiabatic tori orbiting
supermassive SdS black holes ($10^8$--$10^{9}M_\odot$), for which the
test-fluid approximation is valid, are in the case of adiabatic index
$\gamma=7/5$ one or two orders higher than the typical density of molecular
clouds $10^{-18}{\rm kg\,m}^{-3}$ \cite{Car-Ost:1996:ModAst:}.
Of course, to get a more realistic picture, influence of the cosmic repulsion
on self-gravitating tori has to be studied. 
On the other hand, since the jets escaping from AGN can many times exceed the
dimension of the galaxy, repulsive cosmological constant could play an
important role in the collimation of jets far away from the ``seed'' galaxy.

%%%%%%%%%%%%%%%%%%%%%%%%%%%%%%%%%%%%%%%%%%%%%%%%
%% BACKMATTER
%%%%%%%%%%%%%%%%%%%%%%%%%%%%%%%%%%%%%%%%%%%%%%%%

\begin{theacknowledgments}
This work was supported by the Czech grant MSM~4781305903.
\end{theacknowledgments}


\bibliographystyle{aipproc}   % if natbib is available

%\bibliography{mrabbrev,abbs,pubs,refs,pros,petr}

\IfFileExists{\jobname.bbl}{}
 {\typeout{}
  \typeout{******************************************}
  \typeout{** Please run "bibtex \jobname" to optain}
  \typeout{** the bibliography and then re-run LaTeX}
  \typeout{** twice to fix the references!}
  \typeout{******************************************}
  \typeout{}
 }

\end{document}